\input{psfig}
\documentstyle[onecolumn,doublespacing]{mn}

\newcommand{\be}{\begin{equation}}
\newcommand{\ee}{\end{equation}}
\newcommand{\bea}{\begin{eqnarray}}
\newcommand{\eea}{\end{eqnarray}}
\newcommand{\bc}{\begin{center}}

\newcommand{\ec}{\end{center}}
\def\spose#1{\hbox to 0pt{#1\hss}}
\newcommand{\lta}{\mathrel{\spose{\lower 3pt\hbox{$\mathchar"218$}}
     \raise 2.0pt\hbox{$\mathchar"13C$}}}
\newcommand{\gta}{\mathrel{\spose{\lower 3pt\hbox{$\mathchar"218$}}
     \raise 2.0pt\hbox{$\mathchar"13E$}}}

\def\H0{$H_0= 100~h~$km\,s$^{-1}$\,Mpc$^{-1}$}

\newif\ifAMStwofonts

\ifoldfss
  \ifCUPmtlplainloaded \else
    \NewTextAlphabet{textbfit} {cmbxti10} {}
    \NewTextAlphabet{textbfss} {cmssbx10} {}
    \NewMathAlphabet{mathbfit} {cmbxti10} {} 
    \NewMathAlphabet{mathbfss} {cmssbx10} {} 
  \fi
  \ifAMStwofonts
    \ifCUPmtlplainloaded \else
      \NewSymbolFont{upmath} {eurm10}
      \NewSymbolFont{AMSa} {msam10}
      \NewMathSymbol{\upi}     {0}{upmath}{19}
      \NewMathSymbol{\umu}     {0}{upmath}{16}
      \NewMathSymbol{\upartial}{0}{upmath}{40}
      \NewMathSymbol{\leqslant}{3}{AMSa}{36}
      \NewMathSymbol{\geqslant}{3}{AMSa}{3E}

      \let\leq=\leqslant \let\le=\leqslant
      \let\geq=\geqslant \let\ge=\geqslant
    \fi
  \fi
\fi 

\ifnfssone
  \newmathalphabet{\mathit}
  \addtoversion{normal}{\mathit}{cmr}{m}{it}
  \addtoversion{bold}{\mathit}{cmr}{bx}{it}
  \newmathalphabet{\mathbfit} 
  \addtoversion{normal}{\mathbfit}{cmr}{bx}{it}
  \addtoversion{bold}{\mathbfit}{cmr}{bx}{it}
  \newmathalphabet{\mathbfss} 
  \addtoversion{normal}{\mathbfss}{cmss}{bx}{n}
  \addtoversion{bold}{\mathbfss}{cmss}{bx}{n}
  \ifAMStwofonts
    \ifCUPmtlplainloaded \else
and
your
      \UseAMStwoboldmath
      \makeatletter
      \new@mathgroup\upmath@group
      \define@mathgroup\mv@normal\upmath@group{eur}{m}{n}
      \define@mathgroup\mv@bold\upmath@group{eur}{b}{n}
      \edef\UPM{\hexnumber\upmath@group}
      \new@mathgroup\amsa@group
      \define@mathgroup\mv@normal\amsa@group{msa}{m}{n}
      \define@mathgroup\mv@bold\amsa@group{msa}{m}{n}
      \edef\AMSa{\hexnumber\amsa@group}
      \makeatother
      \mathchardef\upi="0\UPM19
      \mathchardef\umu="0\UPM16
      \mathchardef\upartial="0\UPM40
      \mathchardef\leqslant="3\AMSa36
      \mathchardef\geqslant="3\AMSa3E

      \let\leq=\leqslant \let\le=\leqslant
      \let\geq=\geqslant \let\ge=\geqslant
    \fi
  \fi
\fi 

\ifnfsstwo
  \DeclareMathAlphabet{\mathbfit}{OT1}{cmr}{bx}{it}
  \SetMathAlphabet\mathbfit{bold}{OT1}{cmr}{bx}{it}
  \DeclareMathAlphabet{\mathbfss}{OT1}{cmss}{bx}{n}
  \SetMathAlphabet\mathbfss{bold}{OT1}{cmss}{bx}{n}

  \ifAMStwofonts
    \ifCUPmtlplainloaded \else
      \DeclareSymbolFont{UPM}{U}{eur}{m}{n}
      \SetSymbolFont{UPM}{bold}{U}{eur}{b}{n}
      \DeclareSymbolFont{AMSa}{U}{msa}{m}{n}
      \DeclareMathSymbol{\upi}{0}{UPM}{"19}
      \DeclareMathSymbol{\umu}{0}{UPM}{"16}
      \DeclareMathSymbol{\upartial}{0}{UPM}{"40}
      \DeclareMathSymbol{\leqslant}{3}{AMSa}{"36}
      \DeclareMathSymbol{\geqslant}{3}{AMSa}{"3E}

      \let\leq=\leqslant \let\le=\leqslant
      \let\geq=\geqslant \let\ge=\geqslant
    \fi
  \fi
\fi 

\ifCUPmtlplainloaded \else
  \ifAMStwofonts \else 
    \def\upi{\pi}
    \def\umu{\mu}
    \def\upartial{\partial}
  \fi
\fi

\title{Migration of giant planets in a time-dependent planetesimal accretion disc.}
\author[A.Del Popolo et al.]
  {A. Del Popolo,$^1$$^,$$^2$$^,$$^3$ and K.Y. Ek\c{s}i$^3$\\
  $^1$ Dipartimento di Matematica, Universit\`{a} Statale di Bergamo,
  Piazza Rosate, 2 - I 24129 Bergamo, ITALY \\
     $^2$ Feza G\"ursey Institute, P.O. Box 6 \c Cengelk\"oy, Istanbul,
     Turkey\\
     $3$  Bo$\breve{g}azi$\c{c}i University, Physics Department,
     80815 Bebek, Istanbul, Turkey
}
\date{Accepted ???
      Received 2000 July 24;
      in original form ???}

\pagerange{\pageref{firstpage}--\pageref{lastpage}} \pubyear{2000}

\begin{document}

\maketitle

\label{firstpage}

\begin{abstract}

In this paper, we further develop the model for the migration of
planets introduced in Del Popolo et al. (2001).
We first model the protoplanetary nebula as a time-dependent
accretion disc and find self-similar solutions to the equations of
the accretion disc that give to us explicit formulas for the
spatial structure and the temporal evolution of the nebula. These
equations are then used to obtain the migration rate of the planet
in the planetesimal disc and to study how the migration rate
depends on the disc mass, on its time evolution and on some values
of the dimensionless viscosity parameter $\alpha$. We find that
planets that are embedded in planetesimal discs, having total mass
of $10^{-4}-0.1 M_{\odot}$, can migrate inward a large distance
for low values of $\alpha$ (e.g., $\alpha \simeq 10^{-3}-10^{-2}$)
and/or large disc mass and can survive only if the inner disc is
truncated or because of tidal interaction with the star. Orbits
with larger $a$ are obtained for smaller value of the disc mass
and/or for larger values of $\alpha$. This model may explain
several orbital features of the recently discovered giant planets
orbiting nearby stars.
\end{abstract}



\section{Introduction}

After the discovery of 51 Peg by Mayor \& Queloz (1995), more than
sixty extrasolar planet candidates have been discovered.
These planets have unexpected properties: three planets (51 Peg,
$\tau$ Boo, $v$ And) are in extremely tight circular orbits with
periods of a few days, two planets ($\rho^1$ Cnc and $\rho$ CrB)
have circular orbits with periods of order tens of days and three
planets with wider orbits (16 Cyg B, 70 Vir and HD 114762) have
very large eccentricities. Between the unexpected properties of
these planets, most of which are Jupiter-mass objects,
particularly noteworthy is the small orbital separations at which
these planets orbit around their parent stars: among the several
tens of planets detected so far, at least fifteen of the planets
orbit at a distance between $\simeq 0.046$ and $0.11 {\rm AU}$
from their parent star. The properties of these planets are
difficult to explain using the quoted standard model for planet
formation (Lissauer 1993; Boss 1995). This standard model predicts
nearly circular planetary orbits, and giant orbital distances
$\geq 1$ AU from the central star so that the temperature in the
protostellar nebula is low enough for icy materials to condense
(Boss 1995, 1996; Wuchterl 1993, 1996).

The most natural explanation for planets on very short orbits is
that these planets have formed further away in the protoplanetary
nebula and they have migrated to the small orbital distances at
which they are observed. Some authors have also proposed scenarios
in which migration and formation were concurrent (Terquem et al.
1999).

So far, four mechanisms have been proposed to explain the presence
of planets at small orbital distances. A first mechanism deals
with dynamical instabilities in a system of giant planets (Rasio
\& Ford 1996; Weidenschilling \& Marzari 1996). The orbits of
planets could become unstable if the orbital radii evolve
secularly at different rates or if the masses increase
significantly as the planets accrete their gaseous envelopes
(Lissauer 1993). In this model, the gravitational interaction
between two planets, during evolution, (Gladman 1993; Chambers et
al. 1996) can give rise to the ejection of one planet, leaving the
other in a smaller orbit. While it is almost certain that this
mechanism operates in many systems with multiple planets, it
cannot account for the relatively large number of
short-period planets observed (Terquem et al. 1999).\\
The second mechanism, called `migration instability' (Murray et
al. 1998), involves a resonant interaction between
the planet and a disc of planetesimals, located in its orbit which
leads to the planetesimals ejection and the inward migration of
the planet. The advantage of this mechanism is that the migration
is halted naturally at short distances when the majority of
perturbed planetesimals collide with the star. Moreover wide
eccentric orbits can also be produced for planets more massive
than $\simeq 3 M_{\rm J}$. However the model has some
disadvantages, since the protoplanetary disc mass required for the
migration of a Jupiter-mass planet to $a \simeq 0.1$ AU is very
large (Ford et al. 1999; Terquem et al. 1999). \footnote{The
migration of a Jupiter mass planet from 5 AU to very small radii
requires about $0.1 M_{\odot}$ in 5 AU.}

\indent The third possible mechanism proposed to explain short
period of the planets is the dissipation in the protostellar
nebula (Goldreich \& Tremaine 1979, 1980; Ward 1986; Lin et al.
1996; Ward 1997). Since, in this model, the time-scale of
migration is $\simeq 10^5 \frac{M_{\rm p}}{M_{\oplus}} {\rm yr}$
(Ward 1997), the migration has to switch off at a critical moment,
if the planet has to stop close to the star without falling in it.
The movement of the planet might be halted by short-range tidal or
magnetic effects from the central star (Lin et al. 1996) (in any
case, as shown by Murray et al. (1998), it is difficult to
explain, by means of these stopping mechanisms, planets with
semi-major axes $a \ge 0.2$ AU).

The fourth mechanism is based upon dynamical friction between the
planet and a planetesimal disc (Del Popolo et al. 2001) \footnote{In contrast 
to the hypothesis proposed by Fernandez \& Ip (1984), migration in the 
model in Del Popolo et al. (2001) does not 
require the presence of asymmetry in the planetesimals distribution}. In that
paper, we showed that dynamical friction between a planet and a
planetesimals disc is an important mechanism for planet migration
and showed that migration of a $1 M_J$ planet to small
heliocentric distances ($0.05 $ AU) is possible for a disc with a
total mass of $10^{-4} \div 10^{-2} M_{\odot}$ if the planetesimal
disc does not dissipate during the planet migration or if the disc
has $M_{\rm D} >0.01 M_{\odot}$ and the planetesimals are
dissipated in $\sim 10^8$ yr. The model predicts that massive
planets can be present at any heliocentric distances for the right
value of disc mass and
time evolution. \\
Some advantages of the model are: \\
1) differently from models based on the density wave theory
(Goldreich \& Tremaine 1980; Ward 1986, 1997), our model:\\
a) does not require a peculiar mechanism to stop the inward
migration (Lin et al. 1996). Planet halt is naturally provided by
the model. \\
b) It can explain planets found at heliocentric distances of $>
0.1$ AU or planets having larger values of eccentricity. \\
c) It
can explain metallicity enhancements observed in stars
having planets in short-period orbits.\\
2) Whereas the model of Murray et al. (1998) has the drawback of
requiring very massive discs (Ford et al. 1999; Terquem et al.
1999), our model shows that radial migration is possible with
modest masses of planetesimals discs and predicts
the right metallicity enhancement.\\
A point that requires improvement in our model is the model used
for the planetesimal disc. In fact, in Del Popolo et al. (2001),
we used \"{O}pik (1976) approximation, assuming
that the surface density in
planetesimals varies as $\Sigma_{\rm s}(r)=~\Sigma_{\odot}(1 {\rm
AU}/r)^{3/2}$, where $\Sigma_{\odot}$, the surface density at 1
AU, is a free parameter.
The model described in Del Popolo et al.(2001) can be improved
using a more reliable model for the disc, and in particular using
a time-dependent accretion disc, since it is widely accepted that
the solar system at early phases in its evolution, is well
described by this kind of structure. Moreover, astronomical
observations of the last decade have led to the conclusion that
discs around young stellar objects, for example T Tauri stars, are
Keplerian accretion discs and that they are ever-changing and
having a limited life-span. In the next section, we shall get
self-similar solutions of the diffusion equation and will build up
analytical explicit formulas for the surface density and the other
physical quantities required to calculate the planet migration,
due to dynamical friction between the planet and the planetesimals
in the disc.

The plan of the paper is the following: in Sect. ~2, we introduce
the disc model used to study radial migration. In Sect. 3, we
review the migration model introduced in Del Popolo et al. (2001).
In Sect. ~4 we show the results that can be drawn from our
calculations and finally the Sect. ~5 is devoted to the
conclusions.

\section{Disc model}

In order to study the formation of planetary systems it is
necessary to study the global evolution of solid material which
constitutes, together with gas, the protoplanetary discs. The idea
that discs have an important role in stars and planet formation is
not a new one: papers by Peek (1942), von Weizs\"{a}ker (1943,1948)
and L\"{u}st (1952) introduced the idea that the solar nebula was
an accretion disc, while the seminal paper by Lynden-Bell \&
Pringle (1974) foreshadowed the present view that discs are
commonly found in early stellar formation (Beckwith et al. 1990).
In fact, protostellar discs around young stellar objects that have
properties similar to that supposed for the solar nebula are
common: between 25 to 75\% of young stellar objects in the Orion
nebula seem to have discs (Prosser et al. 1994; McCaughrean \&
Stauffer 1994) with mass $10^{-3} M_{\odot}<M_{\rm d} <10^{-1}
M_{\odot}$ and size $40 \pm 20$ AU (Beckwith \& Sargent 1996). The
previous quoted evidences had led to a large consensus about the
nebular origin of the Solar System. Moreover, the observations of
circumstellar discs surrounding T Tauri stars support the view of
a disc having a limited life-span and characterized by continuous
changes during its life. This means that models like the
minimum-mass model cannot model properly the solar nebula: in
order to model the spatial and temporal changes of the disc, one
must use the theory of time-dependent accretion disc. While
numerical models of time-dependent accretion disc (see Ruden \&
Lin 1986; Ruden \& Pollack 1991; Reyes-Ruiz \& Stepinski 1995) are
used in the astronomical context, in the cosmogonic one they have
not been widely used, at least till recent. In the past, but also
in several recent works, the origin of the Solar System was based
upon a steady-state model, namely the minimum-mass model
(Weidenschilling 1977; Hayashi 1985). Other papers assumed only
spatial, but not temporal, changes in the nebula and the origin
and evolution of the Solar System was studied by means of
steady-state accretion disc models (see Morfill \& Wood 1989;
Stepinski et al. 1993). For the reasons previously described and
for others stressed by Stepinski (1998), the temporal evolution of
the nebula must be taken into account, and this can be
accomplished by means of the time-dependent accretion disc model.
In the following, we introduce a time-dependent accretion disc
model that shall be used in the next sections to study planets
migration.

Before starting it is useful to divide, as customary, the
evolution of the solar nebula in three stages: \\
a) the formation
stage, in which the nebula is build up by
infalling matter; \\
b) the viscous stage, in which internal torques produce the
redistribution of angular momentum;\\
c) the clearing stage, in which the gaseous component of the
nebula is dispersed.

The equation that are to be solved to calculate the properties of
the nebula are the thin-disc set of equations (Frank et al. 1985;
Stepinski 1998):
\begin{equation}
\Sigma =2H\rho  \label{surfdens}
\end{equation}

\begin{equation}
H=\sqrt{2}\frac{c_{s}}{\Omega _{K}}  \label{vertical}
\end{equation}

\begin{equation}
c_{s}^{2}=\frac{k}{\mu m_{p}}T  \label{sound}
\end{equation}

\begin{equation}
\frac{16\sigma }{3\Sigma \kappa }T^{4}=\frac{9}{4} \nu \Sigma
\Omega _{K}^{2} \label{energy}
\end{equation}

\begin{equation}
\nu =\alpha c_{s}H  \label{viscosity}
\end{equation}

\begin{equation}
\kappa =\kappa _{0}\rho ^{a}T^{b}  \label{opacity}
\end{equation}

\begin{equation}
\frac{\partial \Sigma }{\partial t}=\frac{3}{r}\frac{\partial }{\partial r}%
\left[ r^{1/2}\frac{\partial }{\partial r}(\nu \Sigma
r^{1/2})\right] \label{diffusee}
\end{equation}
where $\rho $ is the volume density, $\Sigma $ is the surface
density, $H$ is the half thickness of the disc, $c_{s}$ is the
sound velocity, $\Omega _{\rm K}=\sqrt{GM_{\ast }/r^{3}}$ is the
Keplerian angular velocity, $T$ is the temperature at the vertical
center of the disc, $\nu $ is the viscosity and $\kappa $ is the
Rosseland mean opacity. $k$ is the Boltzmann's constant, $\mu $ is
the mean molecular weight (assumed to be 2.33 (Ruden \& Pollack
1991)), $m_{\rm p}$ is the mass of the proton, $\sigma $ is the
Stephan-Boltzmann constant and $\alpha $ is the dimensionless
viscosity introduced by Shakura \& Sunyaev (1973). In some papers,
it is assumed $\alpha=0.01$ as a fiducial value for a disc driven
by thermal convection
(Ruden \& Pollack 1991; Reyes-Ruiz \& Stepinski 1995). In the
following, we shall perform the calculations for three values of
$\alpha$: $10^{-3}$, $10^{-2}$ and $10^{-1}$, which span the
range from inefficient to highly efficient turbulent convection.
See that all the radial dependence is carried in $\Omega _{K}$.
The previous equations constitute a complete physical model from
which it is possible to calculate the physical quantities of the
nebula. Similar to Stepinski (1998), we have assumed the piecewise
power-law formula describing the Rosseland mean opacity,
$\kappa=\kappa(\rho, T) $ given in Ruden \& Pollack (1991). The
formula we used for $\nu=\nu(\Sigma, r)$ is that of Reyes-Ruiz and
Stepinski (1995). $\tau_{\rm crit}$ represents the critical
midplane optical depth required for convective viscosity to be
present. The value assumed is that of Ruden \& Pollack (1991),
$\tau_{\rm crit} \simeq 1.78$ (see also Ruden \& Pollack (1991)
and Reyes-Ruiz \& Stepinski (1995) for a discussion concerning the
introduction of the critical optical depth).

In order to calculate the properties of the nebula, we shall take
advantage of the fact that the nonlinear diffusion process is
self-similar. We recall that self-similar solutions in accretion
discs have been studied \emph{in general} by Pringle (1974),
Filipov (1984), Filipov, Lyubarskii
\& Shakura (1987); \emph{in the context of AGN} by Cannizzo, Lee
\& Goodman (1990); \emph{in the context of Dwarf Nova outbursts}
by Mineshige (1991); \emph{in the context of fallback discs around
newborn neutron stars} by Mineshige, Nomoto \& Shigeyama (1993)
and \emph{in the context of Anomalous X-Ray Pulsars} by Perna,
Hernquist \& Narayan (1999).

Self-similar solutions, asymptotically exact for real problems,
give very important information about the nature of viscously
evolving discs, which does not follow directly from the numerical
calculations. On the other hand, self-similar disc solutions have
the unphysical property that the inner edge of the disc (inner
radius) $R_{in}$ is at the origin ($R_{in}=0$), not at the surface
of the accreting star or somewhere above, like the magnetic
radius.

When the viscous stage starts, namely  when $\Sigma(r,t)$
evolution is governed by equation (\ref{diffusee}), the nebula has
the initial distribution left by the formation stage processes.
The solution of the nonlinear diffusion equation (equation
(\ref{diffusee})) becomes self-similar (Ruden 1993) after the
short transitional time needed to accommodate the initial
conditions and then the self-similar evolution can describe the
nebula evolution except the transition phase. The latter is only
important for the link it provides to initial conditions. Since
the conditions at the beginning of the viscous stage are poorly
constrained, it is convenient, as remarked by Stepinski (1998), to
assume as initial conditions the distribution of surface density
at the beginning of the viscous regime. \footnote{The specific
form of initial conditions does not much influence the evolution
of the gas, inasmuch as the process is diffusive in nature and the
initial conditions are forgotten after a time short in comparison
with the evolutionary time scales.}
 In order to get the
solution for $\Sigma(r,t)$, we first solve the first five
equations algebraically. As an
intermediate step towards this we solve the equations (\ref{surfdens}),(\ref{vertical}),(\ref%
{sound}),(\ref{viscosity}) and (\ref{opacity}) in terms of $\Omega _{K}$, $%
\Sigma $ and $T$:$\allowbreak $

\begin{eqnarray*}
c_{s} &=&\sqrt{\frac{k}{\mu m_{p}}}T^{\frac{1}{2}} \\
H &=&\sqrt{2}\sqrt{\frac{k}{\mu m_{p}}}\frac{1}{\Omega
_{K}}T^{\frac{1}{2}}
\\
\nu &=&\sqrt{2}\alpha \frac{k}{\mu m_{p}}\frac{1}{\Omega _{K}}T \\
\rho &=&\left( \frac{8k}{\mu m_{p}}\right) ^{-\frac{1}{2}}\Sigma
\Omega
_{K}T^{-\frac{1}{2}} \\
\kappa &=&\kappa _{0}\left( \frac{8k}{\mu m_{p}}\right) ^{-\frac{a}{2}%
}\Sigma ^{a}\Omega _{K}^{a}T^{b-\frac{a}{2}}
\end{eqnarray*}%
Then we place these results into equation (\ref{energy}) and solve
for $T$ in terms of $\Omega _{K}$ and $\Sigma $:

\begin{equation}
T=\left( 27\times 2^{-\frac{11+3a}{2}}\frac{\alpha \kappa _{0}}{\sigma }%
\right) ^{\frac{2}{6-2b+a}}\left( \frac{k}{\mu m_{p}}\right) ^{\frac{2-a}{%
6-2b+a}}\Sigma ^{\frac{2\left( a+2\right) }{6-2b+a}}\Omega _{K}^{\frac{%
2\left( a+1\right) }{6-2b+a}}  \label{temper1}
\end{equation}%
Now we place this into the equation for $\nu $ and obtain it in terms of $%
\Omega _{K}$ and $\Sigma $:%
\[
\nu =\sqrt{2}\alpha ^{\frac{2}{6-2b+a}+1}\left( 27\times 2^{-\frac{11+3a}{2}}%
\frac{\kappa _{0}}{\sigma }\right) ^{\frac{2}{6-2b+a}}\left(
\frac{k}{\mu
m_{p}}\right) ^{\frac{2-a}{6-2b+a}+1}\Sigma ^{\frac{2\left( a+2\right) }{%
6-2b+a}}\Omega _{K}^{\frac{2\left( a+1\right) }{6-2b+a}-1}
\]%
Now using $\Omega _{K}=\sqrt{GM_{\ast }/r^{3}}$ we obtain the
viscosity in
terms of $r$ and $\Sigma $:%
\[
\nu =\sqrt{2}\alpha ^{\frac{2}{6-2b+a}+1}\left( 27\times 2^{-\frac{11+3a}{2}}%
\frac{\kappa _{0}}{\sigma }\right) ^{\frac{2}{6-2b+a}}\left(
\frac{k}{\mu
m_{p}}\right) ^{\frac{2-a}{6-2b+a}+1}\left( GM\right) ^{\frac{a+1}{6-2b+a}-%
\frac{1}{2}}r^{-3\left( \frac{a+1}{6-2b+a}-\frac{1}{2}\right) }\Sigma ^{%
\frac{2\left( a+2\right) }{6-2b+a}}
\]%
which is in the form
\[
\nu =Cr^{p}\Sigma ^{q}
\]%
where
\begin{eqnarray*}
C &=&\sqrt{2}\alpha ^{\frac{2}{6-2b+a}+1}\left( 27\times 2^{-\frac{11+3a}{2}}%
\frac{\kappa _{0}}{\sigma }\right) ^{\frac{2}{6-2b+a}}\left(
\frac{k}{\mu
m_{p}}\right) ^{\frac{2-a}{6-2b+a}+1}\left( GM\right) ^{\frac{a+1}{6-2b+a}-%
\frac{1}{2}} \\
p &=&-3\left( \frac{a+1}{6-2b+a}-\frac{1}{2}\right)  \\
q &=&\frac{2\left( a+2\right) }{6-2b+a}
\end{eqnarray*}%
The solution of the equation (\ref{diffusee}) for such a form of
the viscosity can be obtained following Pringle (1974), and Mineshige et al. (1993), and is given by:
\begin{equation}
\frac{\Sigma }{\Sigma _{0}}=K\left( \frac{t}{t_{0}}\right) ^{\frac{-5}{%
5q-2p+4}}\left( \frac{r}{R(t)}\right) ^{-\frac{p}{q+1}}\left[ 1-\left( \frac{%
r}{R(t)}\right) ^{\frac{2q-p+2}{q+1}}\right] ^{\frac{1}{q}}
\label{solutionn}
\end{equation}
where
\[
K=\left( \frac{2q}{(5q-2p+4)(2q-p+2)}\right) ^{\frac{1}{q}}
\]%
and the outer radius of the disc, $R(t)$, is:
\begin{equation}
R(t)=r_{0}\left( \frac{t}{t_{0}}\right) ^{\frac{2}{5q-2p+4}}
\label{eq:rt}
\end{equation}
The scales $r_{0}$, $t_{0}$ and $\Sigma _{0} $ are free except
that they should satisfy the equation:
\begin{equation}
t_{0}=\frac{1}{3C}r_{0}^{2-p}\Sigma _{0}^{-q}.  \label{scale44}
\end{equation}
Note that for $q=0$ the diffusion equation is linear and the
solution is as follows:
\begin{equation}
\Sigma \left( r,t\right) =\frac{M_{\rm d}\left( 0\right) }{2\pi }\frac{1}{%
3lCa^{2-l}t}\left( \frac{a}{r}\right) ^{\frac{9-4l}{4}}e^{-\frac{r^{l}+a^{l}%
}{3l^{2}Ct}}\times I_{\nu _{1}}\left( \frac{2r^{\frac{l}{2}}a^{\frac{l}{2}}}{%
3l^{2}Ct}\right)   \label{sln}
\end{equation}
where $M_{\rm d}(0)$ is the initial mass of the disc, $l=2-p$,
$\nu _{1}=\frac{1}{4-2p}$ and $I_{\nu _{1}}$ is the modified
Bessel function of order $\nu _{1}$. Here the initial form of
$\Sigma $ is chosen to be a Dirac-delta distribution at radius
$a$.
\[
\Sigma \left( r,t\right) =M_{\rm d}\left( 0\right) \frac{5}{24\pi Ca^{\frac{6}{5}%
}t}\left( \frac{a}{r}\right) ^{\frac{29}{20}}e^{-\frac{r^{\frac{4}{5}}+a^{%
\frac{4}{5}}}{\frac{48}{25}Ct}}\times I_{\nu _{1}}\left( \frac{25r^{\frac{2}{%
5}}a^{\frac{2}{5}}}{24Ct}\right)
\]

The solution for $\Sigma(r,t)$ (equation (\ref{solutionn})) has an
explicit radial dependence while the temporal dependence comes
from the time dependence of the outer radius. Three scaling
parameters are present: $r_0$, $\Sigma_0$, and $t_0$, and this
last implicitly depends on the viscosity. Note that the similarity
solutions have four parameters: the initial disc mass, $M_{\rm
d}$, the initial disc characteristic radius, $R_{\rm in}$, the
value of the viscosity and its radial dependence. The
solutions must be continuous at the boundaries between different
viscosity regimes. We indicate with $r_{\rm ij}$ the boundary
between the i-th and j-th viscosity regime. In order to calculate
the boundaries, $r_{\rm ij}$, and impose that $\Sigma(r,t)_{\rm
i}$ are continuous on them, we may follow the same technique used
by Stepinski (1998). Namely, we write the surface density in the
i-th regime as:
\begin{equation}
\Sigma_{\rm i}^{\ast}(r,t)= F_{0,i} \times \Sigma_{\rm i}(r,t)
\label{}
\end{equation}
where $F_{0,i}$ are constants. At the boundary between the 2 and
the 3 regime the temperature is 150 K. We can substitute this
temperature to the equation for the temperature in the regime 2,
$T_2$, to obtain an equation with two unknowns, $r_{23}$ and the
factor $F_{0,2}$, which comes from the dependency of the
temperature on $\Sigma^{\ast}$. Assuming as normalization
$F_{0,2}=1$, $r_{23}$ can be calculated solving the equation
$T_2(r,t)=150 {\rm K}$. Once the value of $r_{23}$ is
known, we can calculate $F_{0,3}$ in two ways: \\
1) using the equation for the temperature for the 3 regime,
substituting in it $r=r_{23}$ and solving
$T_3(r_{23},F_{0,3},t)=150$ for
$F_{0,3}$.\\
2) Imposing the condition
$\Sigma_2(r_{23})=\Sigma_3(r_{23},F_{0,3})$ and solving for
$F_{0,3}$.\\
This procedure can be repeated for regimes 4 and 5.
To find the $r_{12}$ boundary, we follow the same procedure previously
used but with optical depth rather than temperature, namely we use the
equation for the opacity, $\tau=\frac{\kappa \Sigma}{2}$
(Ruden \& Pollack 1991).
%
%
In this way, it is possible to obtain the boundaries between
opacity regimes, in the same way as Stepinski (1998) (see their
equation (15),(16)). Another way of obtaining approximated values
for the boundaries is that of approximating the temperature using
its dependence in the SIGOR regime and solving the same equations.
%

\subsubsection{Different Opacity Regimes}

Different opacity regimes with different $\kappa _{0}$, $a$ and
$b$ will give different sets of $C$, $p$ and $q$. In Table 1, we
have summarized the opacity and viscosity regimes. As in Stepinski
(1998), we have used the following opacity regimes: MOTOR
(Marginally optically thick opacity regime) (is a subset of IGOR
but is based on optical depth instead of temperature); IGOR (Ice
grains opacity regime); IGSOR (Ice grains sublimate opacity
regime); SIGOR (Silicate and iron grains opacity regime); SIGSOR
(Silicate and iron grains sublimate opacity regime).
\begin{table}
\caption{Summary of opacity and viscosity regimes}
\begin{tabular}{lllllll}
\hline Number & Regime & Applicability & Opacity $\left[
\frac{m^{2}}{kg}\right] $
& $C$ & $p$ & $q$ \\
\hline 1 & MOTOR & $\tau <\tau _{crit}$ & $\kappa =2\times 10^{-5}T^{2}$ & $%
0.14\alpha ^{\frac{6}{5}}\mu ^{-\frac{6}{5}}\tau _{cr}^{\frac{2}{5}}m^{-%
\frac{2}{5}}$ & $\frac{6}{5}$ & $0$ \\
\hline 2 & IGOR & $T<150$ K & $\kappa =2\times 10^{-5}T^{2}$ &
$2.\,\allowbreak
03\times 10^{10}\alpha ^{2}\mu ^{-2}$ & $0$ & $2$ \\
\hline 3 & IGSOR & $150$ K$\leq T<180$ K & $\kappa =1.15\times 10^{17}T^{-8}$ & $%
\allowbreak 2.\,\allowbreak 93\times 10^{-3}\alpha ^{\frac{12}{11}}\mu ^{-%
\frac{12}{11}}m^{-\frac{5}{11}}$ & $\frac{15}{11}$ & $\frac{2}{11}$ \\
\hline 4 & SIGOR & $180$ K$\leq T<1380$ K & $\kappa =2.13\times 10^{-3}T^{\frac{3}{4%
}}$ & $142.\alpha ^{\frac{13}{9}}\mu ^{-\frac{13}{9}}m^{-\frac{5}{18}}$ & $%
\frac{5}{6}$ & $\frac{8}{9}$ \\
\hline 5 & SIGSOR & $T\geq 1380$ K & $\kappa =4.38\times 10^{43}T^{-14}$ & $%
\allowbreak 6.\,\allowbreak 44\times 10^{-3}\alpha ^{\frac{18}{17}}\mu ^{-%
\frac{18}{17}}m^{-\frac{8}{17}}$ & $\frac{24}{17}$ &
$\frac{2}{17}$\\
\hline Note: $m=M_{\ast}/M_{\odot}$
\end{tabular}
\end{table}
%
%
In the case of MOTOR, since $q=0$ the diffusion equation is linear
and finding a solution, which can be expressed in terms of Bessel
functions, is much easier (see equation (\ref{sln})).
%
%
It is interesting to
note the role of opacity in changing the disc properties. For
example, an increase in opacity forces the disc to become
convective much nearer the surface, or in other terms produces a
thickening of the convective zone and an increase in the
convective velocity. This produces an increase in the viscosity
leads to a final decrease in the surface density.

\subsection{Determining The Scale Constants}

Integrating the solution, equation (\ref{solutionn}), gives the
mass of the disc as a
function of time%
\begin{equation}
M_{d}=\int_{0}^{R_{out}}2\pi r\cdot \Sigma dr=K^{q+1}(5q-2p+4)\pi
r_{0}^{2}\Sigma _{0}\left( \frac{t}{t_{0}}\right)
^{-\frac{1}{5q-2p+4}}   \label{Mdisk}
\end{equation}
Let $T_{d}$ be the time when the all
effects due to the initial conditions had vanished and the epoch
of self-similar evolution has started. $T_{d}$ is of the order of
dynamical timescale at the inner radius, $R_{\rm in}$ \footnote{In the following we choose 
$R_{\rm in} =0.037 {\rm AU}$.}:
\[
T_{d}\sim \frac{1}{\Omega _{K}(R_{\rm in})}
\]%
If  the mass of the disc at $T_{d}$ is $M_{d}^{0}$, then
\[
M_{d}^{0}\equiv M_{d}(T_{d})=K^{q+1}(5q-2p+4)\pi r_{0}^{2}\Sigma
_{0}\left( \frac{T_{d}}{t_{0}}\right) ^{-\frac{1}{5q-2p+4}}
\]
Using
equation (\ref{scale44}) in this equation, and solving for
$\Sigma_{0}$, we get:
\begin{equation}
\Sigma _{0}=Ar_{0}^{-\frac{5}{2}}  \label{scale55}
\end{equation}%
where
\begin{equation}
A=\left( \frac{M_{d}^{0}}{(5q-2p+4)\pi K^{q+1}}\right) ^{\frac{5q-2p+4}{%
4q-2p+4}}\left( 3CT_{d}\right) ^{\frac{1}{4q-2p+4}}  \label{A}
\end{equation}%
Using equation (\ref{scale55}) in equation (\ref{scale44}), one
obtains
\begin{equation}
t_{0}=\frac{1}{3C}A^{-q}r_{0}^{\frac{5q+4-2p}{2}}  \label{scale6}
\end{equation}%
Using this last in equation (\ref{eq:rt}), we get:
\[
R(t)=\left( \frac{1}{3C}A^{-q}\right) ^{-\frac{2}{5q-2p+4}}t^{\frac{2}{%
5q-2p+4}}
\]%
So $R(t)$ and thus the solutions are independent of the numerical parameter $%
r_{0}$. So for simplicity we choose
\[
r_{0}=1
\]%
Then we obtain
\[
t_{0}=\frac{1}{3CA^{q}}
\]%
\[
\Sigma _{0}=A
\]
The previous calculations are summarized in Table 2 for different
opacity regimes, namely the table gives information on the scale
parameters $r_0$, $t_0$, and $\Sigma_0$.
\begin{table}
\caption{Scale constants in different opacity regimes}
\begin{tabular}{|l|l|l|l|l|}
\hline Number & Regime & $r_{0}$ & $t_{0}$ & $\Sigma _{0}$ \\
\hline 1 & MOTOR & $1$ & $2.38\alpha ^{-\frac{6}{5}}\mu ^{\frac{6}{5}}\tau _{cr}^{-%
\frac{2}{5}}m^{\frac{2}{5}}r_{0}^{\frac{4}{5}}$ & $\left( \frac{5M_{d}^{0}}{%
8\pi K}\right) \left( 3CT_{d}\right) ^{\frac{5}{8}}$ \\ \hline 2 &
IGOR & $1$ & $1.\,\allowbreak 64\times 10^{-11}\alpha ^{-2}\mu
^{2}\left( \frac{M_{d}^{0}}{14\pi K^{3}}\right)
^{-\frac{7}{3}}\left( 3CT_{d}\right) ^{-\frac{1}{6}}r_{0}^{7}$ &
$\left( \frac{M_{d}^{0}}{14\pi K^{3}}\right) ^{\frac{7}{6}}\left(
3CT_{d}\right) ^{\frac{1}{12}}$ \\ \hline
3 & IGSOR & $1$ & $114.\alpha ^{-\frac{12}{11}}\mu ^{\frac{12}{11}}m^{\frac{5%
}{11}}\left( \frac{11M_{d}^{0}}{24\pi K^{\frac{13}{11}}}\right) ^{-\frac{24}{%
121}}\left( 3CT_{d}\right) ^{-\frac{1}{11}}r_{0}^{\frac{12}{11}}$ & $%
\allowbreak \left( \frac{11M_{d}^{0}}{24\pi K^{q+1}}\right) ^{\frac{12}{11}%
}\left( 3CT_{d}\right) ^{\frac{1}{2}}$ \\ \hline
4 & SIGOR & $1$ & $2.\,\allowbreak 35\times 10^{-3}\alpha ^{-\frac{13}{9}%
}\mu ^{\frac{13}{9}}m^{\frac{5}{18}}\left( \frac{9M_{d}^{0}}{61\pi K^{\frac{%
17}{9}}}\right) ^{-\frac{488}{477}}\left( 3CT_{d}\right) ^{-\frac{8}{53}%
}r_{0}^{\frac{61}{18}}$ & $\left( \frac{9M_{d}^{0}}{61\pi K^{\frac{17}{9}}}%
\right) ^{\frac{61}{53}}\left( 3CT_{d}\right) ^{\frac{9}{53}}$ \\
\hline
5 & SIGSOR & $1$ & $51.\,\allowbreak 8\alpha ^{-\frac{18}{17}}\mu ^{\frac{18%
}{17}}m^{\frac{8}{17}}\left( \frac{17M_{d}^{0}}{30\pi K^{\frac{19}{17}}}%
\right) ^{-\frac{15}{119}}\left( 3CT_{d}\right) ^{-\frac{1}{14}}r_{0}^{\frac{%
15}{17}}$ & $\allowbreak \left( \frac{17M_{d}^{0}}{30\pi K^{\frac{19}{17}}}%
\right) ^{\frac{15}{14}}\left( 3CT_{d}\right) ^{\frac{17}{28}}$ \\
\hline
\end{tabular}
\end{table}
Using $\Omega _{K}=\sqrt{GM_{\ast }/r^{3}}$ in equation
(\ref{temper1})
temperature distribution can be written as
$\allowbreak \,\allowbreak $%
\begin{equation}
T=\left( 27\times 2^{-\frac{11+3a}{2}}\frac{\alpha \kappa _{0}}{\sigma }%
\right) ^{\frac{2}{6-2b+a}}\left( \frac{k}{\mu m_{p}}\right) ^{\frac{2-a}{%
6-2b+a}}\left( GM_{\ast }\right) ^{\frac{\left( a+1\right) }{6-2b+a}}r^{%
\frac{-3\left( a+1\right) }{6-2b+a}}\Sigma ^{\frac{2\left( a+2\right) }{%
6-2b+a}}  \label{temper2}
\end{equation}%
%
%
Equation (\ref{temper2}) displays the radial dependence of the
solutions, the temporal one in terms of the outer radius and the
parameters $\Sigma_0$, $t_0$, $\alpha$ and $\mu$. As in the case
of $\Sigma(r,t)$, the temperature is continuous at $r_{\rm ij}$.

\section{Review of the planets migration model}

\indent The model used to study the migration of extra-solar
planets was introduced in two recent papers (Del Popolo et al.
1999, Del Popolo et al. 2001).
The same model shall be used in the present paper to study the
radial migration of extrasolar planets. The difference between the
present paper and the previous one, regarding the migration of
planets in planetesimal discs (Del Popolo et al. 2001), is only
due to the model used to describe the spatial structure and
temporal evolution of the disc. In the following, we review the
planets migration model introduced in Del Popolo et al.
(1999,2001), which shall be applied to study the planets migration
inside a time-dependent accretion disc. Since the model has
already been described in the two quoted papers, the reader is
referred to those for further details.

We consider a thin planetesimal disc around a star of mass
$M_{\ast}=1 M_{\odot}$ and suppose that a single planet moves in
the disc under the influence of the gravitational force of the
star. The equation of motion of the planet can be written as:
\begin{equation}
{\bf \ddot r}= {\bf F}_{\odot}
+{\bf R}
\end{equation}
(Melita \& Woolfson 1996), where the term ${\bf F}_{\odot}$
represents the force per unit mass from the Sun, while ${\bf R}$
is the dissipative force (the dynamical friction term-see Melita
\& Woolfson 1996). In order to take into account dynamical
friction, we need a suitable formula for a disc-like structure
such as the protoplanetary disc.

We assume that the matter-distribution is disc-shaped and that it
has a velocity distribution described by:
\begin{equation}
n({\bf v},{\bf x})=n({\bf x})\left( \frac 1{2\pi }\right)
^{3/2}\exp \left[ -\left( \frac{v_{\parallel }^2}{2\sigma
_{\parallel
}^2}+%
\frac{v_{\perp }^2}{2\sigma _{\perp }^2}\right) \right] \frac
1{\sigma _{\parallel }^2\sigma _{\perp }}
\end{equation}
(Hornung \& al. 1985, Stewart \& Wetherill 1988) where $
v_{\parallel }$ and $\sigma_{\parallel}$ are the velocity and the
velocity dispersion in the direction parallel to the plane while $
v_{\perp }$ and $\sigma_{\perp}$ are those in the perpendicular
direction. We suppose that  $\sigma_{\parallel}$ and
$\sigma_{\perp}$ are constants and that their ratio is simply
taken to be 2:1 ( $\sigma_{\parallel}$=$2 \sigma_{\perp}$).

Then according to Chandrasekhar (1968) and Binney (1977) we may
write the force components as:
\begin{eqnarray}
F_{\parallel } &= & k_{\parallel}v_{1 \parallel}= B_{\parallel
}v_{1\parallel } \left[ 2\sqrt{2\pi } \overline{n} G^2\log \Lambda
m_1m_2\left( m_1+m_2\right) \frac {\sqrt{1-e^2}}{\sigma
_{\parallel }^2\sigma _{\perp }}\right]\label{eq:b1}
\end{eqnarray}
\begin{eqnarray}
F_{\perp } &= &k_{\perp}v_{1 \perp}= B_{\perp }v_{1_{\perp }}
\left[ 2\sqrt{2\pi } \overline{n} G^2\log \Lambda m_1m_2\left(
m_1+m_2\right) \frac {\sqrt{1-e^2}}{\sigma _{\parallel }^2\sigma
_{\perp }}\right]\label{eq:b2}
\end{eqnarray}
where
\begin{eqnarray}
B_{\parallel } &= &\int_0^\infty \exp{ \left[ -\frac{v_{1\parallel
}^2}{%
2\sigma _{\parallel }^2}\frac 1{1+q}-\frac{v_{1\perp }^2}{2\sigma
_{\parallel }^2}\frac 1{1-e^2+q}\right] } \times
\frac {dq}{\left[ \left( 1+q\right) ^2\left( 1-e^2+q\right)
^{1/2}\right]} \label{eq:b3}
\end{eqnarray}
\begin{eqnarray}
B_{\perp } &= &\int_0^\infty \exp \left[ -\frac{v_{1\parallel
}^2}{2\sigma _{\parallel }^2}\frac 1{1+q}-\frac{v_{1\perp
}^2}{2\sigma _{\parallel
}^2}%
\frac 1{1-e^2+q}\right]  \times
\frac {dq}{\left[ \left( 1+q\right) \left( 1-e^2+q\right)
^{3/2}\right] } \label{eq:b4}
\end{eqnarray}
and
\begin{equation}
e=(1-\sigma_{\perp}^2/\sigma_{\parallel}^2)^{0.5}
\end{equation}
while $\overline{n}$ is the average spatial density, $m_1$ is the
mass of the test particle, $m_2$ is the mass of a field one, and
$\log{\Lambda}$ is the Coulomb logarithm. The frictional drag on
the test particles may be written as:
\begin{equation}
{\bf F}=-k_{\parallel}v_{1 \parallel} {\bf e_{\parallel}}-
k_{\perp}v_{1 \perp}{\bf e_{\perp}} \label{eq:dyn}
\end{equation}
where ${\bf e_{\parallel}}$ and ${\bf e_{\perp}}$ are two unit
vectors parallel and perpendicular to the disc plane.\\
Since damping of eccentricity and inclination is more rapid than
radial migration (Ida 1990; Ida \& Makino 1992; Del Popolo et al.
1999), we deal only with radial migration and we assume that the
planet has negligible inclination and eccentricity, $i_{\rm p}
\sim e_{\rm p} \sim 0$ and that the initial heliocentric distance
of the planet is $5.2$ ${\rm AU}$. For the objects lying in the
plane, the dynamical drag is directed in the direction opposite to
the motion of the particle and is given by (Del Popolo et al.
1999):
\begin{equation}
{\bf F} \simeq -k_{\parallel}v_{\parallel} {\bf e_{\parallel}}
\end{equation}

In the simulation, we assume that the planetesimals all have
equal masses, $m$, and that $m<< M$, $M$ being the planet mass.
This assumption does not affect the results, since dynamical
friction does not depend on the individual masses of these
particles but on their overall density. If the planetesimals
attain dynamical equilibrium, their equilibrium velocity
dispersion, $\sigma_{\rm m}$, would be comparable to the surface
escape velocity of the dominant bodies (Safronov 1969), while if
we consider a two-component system, consisting of one protoplanet
and many equal-mass planetesimals the velocity dispersion of
planetesimals in the neighborhood of the protoplanet depends on
the mass of the protoplanet. Since the eccentricity is given by: \\
\begin{equation}
e_{\rm m} \simeq \left\{
\begin{array}{lc}
20 (2 m /3 M_{\odot})^{1/3} & M \le 10^{25} g  \\
6 (M/3M_{\odot})^{1/3}  & M > 10^{25} g
\end{array}
\right.
\end{equation}
(Ida \& Makino 1993) where $m$ is the mass of the planetesimals,
%
then the dispersion velocity in the disc is characterized by two
regimes, being it connected to the eccentricity by the equation:
\begin{equation}
\sigma_{\rm m} \simeq (e_{\rm m}^2+i_{\rm m}^2)^{1/2} v_{\rm c}
\end{equation}
where $i_{\rm m}$ is the inclination of planetesimals and $v_{\rm
c}$ is the Keplerian circular velocity.
In the neighborhood of the protoplanet there is a region having a
larger dispersion velocity.
The width of this ``heated"
region is roughly given by $4 [(4/3)(e_{\rm m}^2+i_{\rm
m}^2)a^2+12 h_{\rm M}^2 a^2]^{1/2}$ (Ida \& Makino 1993) where $a$
is the semi-major axis and $h_{\rm M}= (\frac{M+m}{3
M_{\odot}})^{1/3}$ is the Hill radius of the protoplanet. The
increase in velocity dispersion of planetesimals around the
protoplanet decreases the dynamical friction force (see Eq.
\ref{eq:dyn}) and consequently
increases the migration time-scale.\\
In our model the gas is almost totally dissipated when the planet
begins to migrate.
We assume that the spatial structure and the time evolution of the
surface density, $\Sigma_{\rm s}(R,t)$, in planetesimals is described by
the disc model introduced in the previous section with the further 
assumption that 
%
%
the initial radial distribution of solids is $\Sigma_{\rm s}(t=0,r)=\Sigma(t=0,r) \delta$, where $\delta=10^{-2}$ for ice (gas-giant region) and $\delta=6 \times 10^{-3}$ for silicates (terrestrial planet region), to account for cosmic abundance.

The total mass
in the planetesimal disc, for a fixed viscosity regime, is then:
\begin{eqnarray}
M_d(t) &=&\int\limits_0^R\Sigma \cdot 2\pi rdr  \nonumber \\
&=&M_0\left( \frac t{t_0}\right) ^{-\frac 1{5q-2p+4}}
\label{Mdisk}
\end{eqnarray}
 We integrated the equations of motion using the Bulirsch-Stoer
method.

Before going on it is important to discuss a clue assumption of the paper, above introduced, namely
that the surface density in planetesimals is proportional to that of gas
and that the spatial structure and the time evolution of the
surface density, $\Sigma_{\rm s}(R,t)$, is such that $\Sigma_{\rm s}(R,t) \propto \Sigma(R,t)$.

We know that the evolution of the surface density of gas, Eq. (\ref{diffusee}), is described 
by a diffusive-type equation, while that of solid particles is an advection-diffusion equation:
\begin{equation}
\frac{\partial \Sigma_{\rm s} }{\partial t}=\frac{3}{r}\frac{\partial }{\partial r}%
\left[ r^{1/2}\frac{\partial }{\partial r}(\nu_{\rm s} \Sigma_{\rm s}
r^{1/2})\right]+
\frac{1}{r}\frac{\partial }{\partial r}\left[ \frac{2r
\Sigma _{s}\langle \overline{v}_{\phi }\rangle _{s}}{\Omega _{k}t_{s}}\right]  
\label{advdiffusee}
\end{equation}
where $\nu_{\rm s}=\frac{\nu}{{\rm Sc}}$, the Schmidt number, Sc, is given by:
\begin{equation}
{\rm Sc}=\left( 1+\Omega _{\rm k}t_{\rm s}\right) \sqrt{1+\frac{\overline{\bf v}^2}{V_{\rm t}^2}}
\end{equation}
where ${\bf v}$ is the relative velocity between a particle and the gas, $V_{\rm t}$ the 
turbulent velocity, $\Omega _{\rm k}$ is the Keplerian angular velocity, $t_{\rm s}$ the so called stopping time.
If $\Omega _{\rm k}t_{\rm s} \rightarrow 0$, the stopping time is small in comparison with the period of orbital revolution and particles are strongly coupled to the gas (particles of size $<1$ mm).
If $\Omega_{\rm k} t_{\rm s} \rightarrow \infty$, the stopping time is very long in comparison with the period of orbital revolution and particles are decoupled from the gas (particles of size $>10^4$ cm).
Then the mass distribution of planetesimals emerging from a turbulent disc does not necessarily reflect that of gas.
In general, the mass distribution of solids evolves due to gas-solid-coupling, coagulation, sedimentation, and evaporation/condensation. After some preliminary studies of Cassen (1996) and Schmitt et al. (1997), Stepinski \& Valageas (1996, 1997) developed a method that, using a series of simplifying assumptions, is
able to simultaneously follow the evolution of gas and solid particles due to gas-solid-coupling, coagulation, sedimentation, and evaporation/condensation for up to $10^7 {\rm yr}$. The model is based on the premise that the  
transformation of solids from dust to planetesimals occurs fundamentally through the process of hierarchical coagulation (other possibility for this transformation are reported in Goldreich \& Ward 1973; Barge \& Sommeria 1995; Tanga et al. 1996). The model is a hypothetical evolutionary scenario that can at least be used to illustrate the difference in 
time evolution between gas and solids (I want to recall that a further element of 
imprecision is the dust opacitiy which may be uncertain in the mm range 
by a factor of 4-5 (Pollack et al. 1994)). Stepinski's model, goes only to times of $10^7$ years fundamentally 
because
at such long times solids are mostly in large bodies (planetesimals)
and Stepinski's method, which neglect gravitational interactions between solids, is
not longer a reasonable approximation. 

The comparison of the evolution of the gas disc and that of solids is plotted in Fig.4-5.   
Fig. 4 shows the evolution of the mass of the nebula
for a disc of $M_{\rm d}=0.245 M_{\odot}$, $\alpha=0.01$.
The solid line represents the surface density of the gas $\times 10^{-2}$ (obtained with our model),
while the dashed line the evolution of long lived particles (characterized by evolutionary 
timescales comparable or longer than those of gas) of $10^4 {\rm cm}$, calculated by Stepinski \& Valageas (1996).

Fig. 5, shows the evolution of the mass of the nebula
for a disc of $M_{\rm d}=0.023 M_{\odot}$, for $\alpha=0.1$ (Fig.5a), $\alpha=0.01$ (Fig. 5b),$\alpha=0.001$ (Fig. 5c),
$\alpha=0.0001$ (Fig. 5d).
The dashed line represents the surface density of the gas $\times 0.01$ (obtained with our model),
while the dot-dashed line the evolution of solids,  calculated by Stepinski \& Valageas (1997).

Summarizing, the previous plots, Fig. 4-5, show that the difference in evolution of gas $\times 10^{-2}$ and that 
of solids becomes smaller for smaller values of $\alpha$. 

Although, as remarked, the evolution of gas and solids is different, I'll use in the present paper the quoted approximation 
$\Sigma_{\rm s}(R,t) \propto \Sigma(R,t)$. There are several reasons for this choice:\\
a) In the past, but also
in several recent works, the study of origin of the Solar System was based
upon a steady-state model, namely the minimum-mass model
(Weidenschilling 1977; Hayashi 1985). Other papers assumed only
spatial, but not temporal, changes in the nebula and the origin
and evolution of the Solar System was studied by means of
steady-state accretion disc models (see Morfill \& Wood 1989;
Stepinski et al. 1993). 
So, the use, in this paper of a time-dependent accretion disc, is surely an improvement on the previous papers.\\
b) The assumption that the surface density of the planetesimals disc is $0.01$ of the gas disc is used by 
several authors: for example, Terquem et al. (2000) (page 3), write:
``the surface mass density of the planetesimal disk was derived from $\Sigma$ (the gas surface density) by noting that in protostellar disks the gas to dust ratio is about 100". This assumption is not ``exceptionally" used by the prevìous authors, but it is almost 
the ``rule" in literature (e.g., Lin \& Papaloizou (1980, page 47); Murray et al. (1998)). The paper of Stepinski \& Valageas (1996), which is the first attempt to study, numerically, the global evolution of solids in a protoplanetary discs, recognizes that the above quoted assumption is notheworthy diffused in literature, and nowaday it is always used.\\

\section{Results}

 In the present paper, we are
fundamentally interested in studying the planets migration due to
interaction with planetesimals and for this reason we suppose that
the gas is almost dissipated when the planet starts its migration.
Clearly the effect of the presence of gas should be that of
accelerating the lose of angular momentum of the planet and to
reduce the migration time. Similarly to Del Popolo et al. (2001),
we assume that the gas disc has a nominal effective lifetime of
$10^6$ years (Zuckerman et al. 1995), compatible with several
evidences showing that the disc lifetimes range from $ 10^5$ yr to
$10^7$ ~yr (Strom et al. 1993; Ruden \& Pollack 1991). Usually,
this decline of gas mass near stars is more rapid than the decline
in the mass of orbiting particulate matter (Zuckerman et al.
1995). Then the disc is populated by residual planetesimals for a
longer period. Summarizing our model starts with a fully formed
gaseous giant planet of $1 M_{\rm J}$ at $5.2$ AU embedded in disc
of planetesimals, without gas. The model introduced in the
previous sections was integrated for several values of the disc
surface density or equivalently several disc
masses: $M_{\rm D}=0.1$, 0.01, 0.005, 0.001, 0.0005, 0.0001 $M_{\odot}$ 
(note that we follow the same notation of Del Popolo et al. (2001) and so the 
value of the disc mass, $M_{\rm D}$, refers to initial gas disc).\\
We recall that estimates of the ``minimum mass" disc necessary to
form our planetary system range from $\simeq 0.001$ to $\simeq 0.1
M_{\odot}$ (Weidenschilling 1977; Boss 1996). Due to the loss of
planetesimals ejected through orbital encounters with giant
planets (see Murray et al. 1998), estimates on the low end of this
range may be insufficient to form our solar system; consequently,
the minimum mass for our solar system may be probably $\simeq 0.06
M_{\odot}$ (Boss 1996). The results of the disc model introduced
in section ~(2) are plotted in Fig.1-3.

Fig. 1 shows the detailed evolution of the surface density and we
compare the surface density obtained with the model of section
(2), solid line, with that of Stepinski (1998), dashed line, for a
disc mass of $0.1 M_{\odot}$, $\alpha=0.01$, $\tau_{\rm
crit}=1.78$ and the angular momentum given by:
\begin{equation}
J_{\rm d}=\int\limits_0^Rr^2\Omega_{\rm K} \Sigma \cdot 2\pi rdr
\end{equation}
The angular momentum can be directly obtained using equation
(\ref{solutionn}), and for a fixed viscosity regime is given by:
\begin{equation}
J_{\rm d}=\left(\frac{2K(q+1) B_1}{2q-p+2}\right) \pi r_0^2\Sigma
_0\sqrt{GMr_0}
\end{equation}
where
\[
B_1=B(\frac{q+1}q,\frac 12\frac{5q+5-2p}{2q-p+2})
\]
is the beta function defined as
\[
B(k,l)=\frac{\Gamma (k)\Gamma (l)}{\Gamma (k+l)}
\]
which for the disc parameter previously defined is 
$4 \times 10^{52} {\rm g cm^2/s }$. The plots, from top to bottom, represent
the surface density at times $t=10^5$, $10^6$, $10^7 {\rm yr}$.
The surface density plotted was obtained by imposing, similar to
Stepinski (1998), the continuity of $\Sigma$ at boundaries between
different viscosity regimes.
The size of the viscosity regimes
and the matching of $\Sigma(r,t)$ at the $r_{ij}$ was obtained as
described in section (2).
%
%
As shown in the figure, our model for $\Sigma$ is
in good agreement with Stepinski's (1998) model. As shown, there
are already five different regimes present at $t=10^5 {\rm yr}$.
In general the angular momentum $J_{\rm d}$ and the disc mass
$M_{\rm d}$ determine the overall evolution of the nebula and
$\alpha$ sets the time scales: lower values of $\alpha$ give rise
to longer time scales. The number of epochs in the nebula
evolution are strictly connected to the initial conditions,
$M_{\rm d}$ and $J_{\rm d}$. In fact, fixing the initial mass,
the smaller the angular momentum, the more concentrated towards
the star is the mass distribution. Usually there are two or more
regimes and consequently different parts of it have a behavior set
by the relative viscosity regime. Differently from other models,
like Dubruelle (1993) and  Cassen (1994,1996), the changes in
$\Sigma$ are not given by a simple decay formula (see Stepinski
1998, equation (1)). This is due to the presence of different
epochs and since the viscosity is described but more than one
power-law. Fig. 1 shows that at $10^5 {\rm yr}$ the disc is in the
main viscous phase. The accretion of mass towards the central star
gives rise to a later evolution characterized by a steady decrease
in the surface density. The evolution proceeds self-similarly.
%
%
In Fig. 2, we also compared the result of our model, solid line, with the
numerical simulations of Reyes-Ruiz \& Stepinski (1995), dashed
line, and $\Sigma$ obtained using our model but only the viscosity
regime SIGOR, dotted line.
As shown there is a good agreement between our analytic model and their
numerical simulations.
%
%
The main differences between the analytical
model and numerical simulations is larger in the innermost regions
of the nebula. This is due to the fact that in numerical
simulations Reyes-Ruiz \& Stepinski (1995) has used additional,
high-temperature opacity regimes, neglected in the analytical
model. It is also interesting to note that the analytical
prediction for $\Sigma$ obtained using only the SIGOR regime is
also in good agreement with the other two curves at least in the
region 0.3-5.2 AU, which is of particular interest for the
migration of giant planets studied in this paper. Fig. 3 plots the
radial distribution of temperature, $T(r,t)$, at selected times
($10^5 {\rm yr}$, $10^6 {\rm yr}$, $10^7 {\rm yr}$) for a disc of
$M_{\rm d}=0.1 M_{\odot}$ and $\alpha=0.01$.
The
solid line represents $T(r,t)$ obtained with Stepinski's (1998)
model, the dashed line the prediction for temperature of the model
of this paper and the dotted line the approximation given by using
only the SIGOR regime. As in the case of $\Sigma(r,t)$, the
temperature obtained in this paper is in good agreement with
Stepinski's (1998) model and Reyes-Ruiz \& Stepinski (1995).

In the successive figures (6, 7, 8), we plot the evolution of
semi-major axis of the planet. In Fig. 6, we plot the evolution of
a 1 $M_{J}$ planet in a disc with planetesimals whose surface
density is 1\% of the gas (Lin \& Papaloizou 1980; Stepinski \&
Valageas 1996) and having $\alpha=0.001$. The disc model used is
that introduced in section 2. As previously reported, the
simulation is started with the planet at 5.2 AU and $i_{\rm p}
\sim e_{\rm p} \sim 0$. The values of the initial disc mass are
$M_{\rm D}$: 0.1, 0.01, 0.005, 0.001, 0.0005, 0.0001 $M_{\odot}$,
for the solid line, dotted line, short-dashed line, long-dashed
line, dot-short dashed line, dot-long dashed line, respectively.
As it is evident, a disc of larger mass produces a more rapid
migration of the planet. Fig. 6 shows that the planet embedded in
a disc having $M_{\rm D}=0.1, 0.01$, 0.005, 0.001, 0.0005, 0.0001
$M_{\odot}$ migrates to 0.05 AU in $9.5 \times 10^7 {\rm yr}$,
$2.7 \times 10^8 {\rm yr}$, $3.7 \times 10^8 {\rm yr}$, $8 \times
10^8 {\rm yr}$, $1.2 \times 10^9 {\rm yr}$, $2.5 \times 10^9 {\rm
yr}$, respectively. It is important to stress that even if the
planet reaches distances of $0.05 {\rm AU}$ in less than $4.5
\times 10^9 {\rm yr}$, the planet must halt at several $R_{\ast}$
from the star surface ($R_{\ast}$ is the stellar radius). In fact
solid bodies cannot condense at distances $\leq 7 R_{\ast}$, and
planetesimals cannot survive for a long time at distances $\leq 2
R_{\ast}$. When the planet arrives at this distance the
dynamical friction force switches off and its migration stops.\\
This means that the minimum value of the semi-major axis that a
planet can reach is $\simeq 0.03$
AU.\\
In order to study the effect of viscosity on migration, we
performed two other simulations with $\alpha=0.01$ and
$\alpha=0.1$, and plotted in Fig. 7 and Fig. 8, respectively. In
Fig. 7, we plot the case $\alpha=0.01$. As for Fig. 6 the values
of the initial disc mass are $M_{\rm D}$: 0.1, 0.01, 0.005, 0.001,
0.0005, 0.0001 $M_{\odot}$, for the solid line, dotted line,
short-dashed line, long-dashed line, dot-short dashed line,
dot-long dashed line, respectively. The plot shows that for
$M_{\rm D}=0.1 M_{\odot}$ the planet moves to $0.05$ AU in $\simeq
8.8 \times 10^8$ yr. If $M_{\rm D}=0.01, 0.005 M_{\odot}$, the
time needed to reach 0.05 AU is given by $\simeq 2.5 \times 10^9$
yr, $\simeq 3.2 \times 10^9$ yr, respectively. If the disc mass is
$M_{\rm D}=0.002 M_{\odot}$, the time needed for the planet to
reach the quoted position is larger than the age of the stellar
system.
Similar to the case for $\alpha=0.001$, even in the cases $M_{\rm
D}=0.1, 0.01, 0.005 M_{\odot}$, in which the planet reaches
distances of $0.05 {\rm AU}$ in less than $4.5 \times 10^9 {\rm
yr}$, the planet must halt at several $R_{\ast}$ from the star
surface ($R_{\ast}$ is the stellar radius).
In the case $M_{\rm D}=0.001 M_{\odot}$, the planet stops its
migration at $a \simeq 0.3$ AU, while if $M_{\rm D}=0.0005,
0.0001, M_{\odot}$, we have $a \simeq 0.7, 1.86$ AU. Fig. 8 is the
same of the previous two figures except that $\alpha=0.1$ in this
case. In particular, we find that the planet embedded in a disc
having $M_{\rm D}=0.1, 0.01$, 0.005, 0.001, 0.0005, 0.0001
$M_{\odot}$ migrates to 0.53, 2.12, 2.57, 3.43, 3.73, 4.25 AU,
respectively, in $4.5 \times 10^9 {\rm yr}$. We see that the
migration time increases going from $\alpha=0.001$ to $\alpha=0.1$
since discs with lower value of $\alpha$ evolve slower than discs
with larger values of this parameter, and the high value of the
gas and particle density is retained for a much longer time
(Stepinski \& Valageas 1996). As the solar nebula evolution is
faster for larger values of $\alpha$, larger amounts of mass are
lost from the nebula, in a given time, and since a less massive
disc implies a less rapid migration, the planet shall move less, in
the case of larger $\alpha$
(see Ruden \& Pollack 1991, Fig.1).
%

The final distribution of planets shows that, in the case of discs
having a value of $\alpha=0.001$, for disc mass in the range $0.1
M_{\odot}<M<0.0001 M_{\odot}$, a Jupiter-like planet in any case
migrates to a very small distance from the parent star and that,
for the reason described above, the migration stops at several
$R_{\ast}$. If the value of $\alpha=0.01$, planets can migrate
both to very short distances from the star (for disc masses in the
range $0.1 M_{\odot}<M<0.005 M_{\odot}$) or to distances $\simeq 2
{\rm AU}$ for $M \simeq 0.0001 M_{\odot}$. Planets cannot migrate
to very small distances from the star, if $\alpha=0.1$ and $0.1
M_{\odot}<M<0.0001 M_{\odot}$. In this last case, the planet is
located in the range $0.5 < a < 4$. Summarizing,
according to the final distribution of planets distances, similar
to Del Popolo et al. (2001), the present model predicts that
planets can be present at any distance from their locations of
formation and very small distances from the parent star.
Differently from Del Popolo et al. (2001), now the location of the
planets also depend on the viscosity through the parameter
$\alpha$. It is evident from the previous results that the
viscosity has at least the same importance of disc mass for what
concerns migration. Another difference that our new disc model has
on  migration of planets is the different time scale and that the
evolution of the semi-major axis never goes to a constant value as
is did in Del Popolo et al. (2001). This last difference is due to
the more rapid decrease in surface density and disc mass assumed
in Del Popolo et al. (2001) (see Fig. 2 of Del Popolo et al.
(2001)). One of the aims of this last paper was that of finding the
qualitative effect of disc evolution and this led us to assume a
decay law for the disc mass similar to that for cm size particles,
given in Stepinski \& Valageas (1996). This was only a rough
assumption, to have an idea of the effect of mass evolution in the
disc on the migration of planets. In the present paper, the disc
model gives us a better background to understand how the spatial
structure and the temporal evolution of the nebula influences the
migration of planets.

The configuration of observed planets can be reproduced by means
of a combination of $\alpha$ and $M_{\rm d}$. For example, the
parameters of 47 UMa b ($a=2.11$ AU) can be explained, for
example, with $\alpha=0.01$ and $M_{\rm d} \simeq 0.0001$ or
$\alpha=0.1$ and $M_{\rm d} \simeq 0.01$, while planets like
$\tau$ Bootis b or 51 Peg b, having very small semi-major axis,
can be reproduced with $\alpha=0.001$ and $0.1 M_{\odot}<M_{\rm
d}<0.0001 M_{\odot}$ or $\alpha=0.01$ and $0.1 M_{\odot}<M_{\rm
d}<0.0035 M_{\odot}$. Configurations similar to that of 55 Cnc b
($a=0.11$ AU), $\rho$ CrB b ($a=0.23$ AU) are obtained for
$\alpha=0.01$, $M_{\rm d} \simeq 0.0015 M_{\odot}$ and
$\alpha=0.01$ $M_{\rm d} \simeq 0.0012 M_{\odot}$, respectively.
The semi-major axis of planets like 70 Vir b ($a=0.43$, $e_{\rm
p}=0.4$; $M \sin{i_{\rm p}} \sim 6.6 M_J$) and HD 114762 b
($a=0.3$; $e_{\rm p}=0.25$; $M \sin{i_{\rm p}} \sim 10 M_{\rm J}$)
can be explained by radial migration, as shown, while their high
value of eccentricities can be explained as described in Del
Popolo et. al (2001). \footnote{If a planet having mass $M \ge 3
M_{\rm J}$ moves in a planetesimal disc during interactions,
planetesimals scattered from their Hill sphere can be ejected with
$|\frac{\Delta E}{\Delta L}|<1$, (where $\Delta E$ and $\Delta L$
are respectively the energy and angular momentum removed from a
planet by the ejection of a planetesimal), and the eccentricity
$e_{\rm p}$ tends to increase.}
For
what concerns the enhancement of metallicity of stars with
short-period planets, having high metallicities, $[{\rm Fe/H}]
\geq 0.2$ (Gonzales (1997; 1998a,b), there is no difference with
Del Popolo et al. (2001); namely we expect that the planet
delivers into the star, in the case of a disc $M_{\rm D}=0.01
M_{\odot}$, a mass of $M_{\rm acc} \sim 40 M_{\oplus}$ which
means $[{\rm Fe/H}] \sim 0.2$.
We can add that, since the quoted
mass is delivered if the planet moves to a distance of $0.05 {\rm
AU }$, the disc should preferably have $\alpha=0.001$ and $M_{\rm
d}=0.01 M_{\odot}$, $\alpha=0.01$ and $M_{\rm d}=0.01 M_{\odot}$
or $\alpha=0.1$ and $M_{\rm d}=0.1 M_{\odot}$ (note that this mass
should be contained in a disc of several tenths of AU).
\section{Conclusions}

In this paper, we have studied the effect of dynamical friction on
the migration of a giant planet in a planetesimal time-dependent
accretion disc, having only $\simeq 1 \%$ of its mass in the form
of solid particles (Stepinski \& Valageas 1996, Murray et. al
1998). The paper is an improvement of a previous one (Del Popolo
et al. 2001), in which the effect of dynamical friction has been
studied for a disc having surface density
$\Sigma(r)=\Sigma_{\odot}(1 {\rm AU}/r)^{3/2}$, where
$\Sigma_{\odot}$, the surface density at 1 AU, is a free
parameter. To start with, we found a self-similar solution to the
diffusion equation: the solution obtained is very similar to that
obtained by Stepinski (1998) in his theoretical model. Our
theoretical model for the disc is in good agreement with numerical
calculations of Reyes-Ruiz \& Stepinski (1995) and with the
theoretical model of Stepinski (1998). The disc model was used to
calculate the planet migration. We found in the case of discs
having a value of $\alpha=0.001$, and for disc mass in the range
$0.1 M_{\odot}<M<0.0001 M_{\odot}$, that a Jupiter-like planet in
any case migrates to a very small distance from the parent star
and that, for the reason described above, the migration stops at
several $R_{\ast}$. If the value of $\alpha=0.01$, planets can
migrate both to very short distances from the star (for disc
masses in the range $0.1 M_{\odot}<M<0.005 M_{\odot}$) or to
distances $\simeq 2 {\rm AU}$ for $M \simeq 0.0001 M_{\odot}$.
Planets cannot migrate to very small distances if $\alpha=0.1$ and
$0.1 M_{\odot}<M<0.0001 M_{\odot}$. In this last case, the planet
is located in the range $0.5 <a <4$. The location
of planets depend on two parameters: the viscosity, through the
parameter $\alpha$, and the disc mass. Different configurations of
observed planets (e.g., 47 UMa b; $\tau$ Bootis b; 51 Peg b; 55
Cnc b; $\rho$ CrB b; 47 UMa b) can be explained by combinations of
$\alpha$, and the disc mass. The configuration of large
eccentricity planets like 70 Vir b ($a=0.43$, $e_{\rm p}=0.4$; $M
\sin{i_{\rm p}} \sim 6.6 M_J$) and HD 114762 b ($a=0.3$; $e_{\rm
p}=0.25$; $M \sin{i_{\rm p}} \sim 10 M_{\rm J}$) can be explained
by radial migration, while their high value of eccentricities can
be explained observing that, in the case $M \ge 3 M_{\rm J}$,
eccentricity the planets $e_{\rm p}$ tends to increase if
planetesimals scattered from their Hill sphere can be ejected with
$|\frac{\Delta E}{\Delta L}|<1$. Similar to Del Popolo et al.
(2001), metallicity enhancement observed in several stars having
extrasolar planets can also be explained by means of scattering of
planetesimals onto the parent star, after the planet reached its
final configuration. However, the disc should preferably have
$\alpha=0.001$ and $M_{\rm d}=0.01 M_{\odot}$, $\alpha=0.01$ and
$M_{\rm d}=0.01 M_{\odot}$ or $\alpha=0.1$ and $M_{\rm d}=0.1
M_{\odot}$.
%
%

Interesting points to address in a future study are:\\
a) migration in presence of at least two planets;\\
b) effect of gas and dynamical friction on the migration; \\
c) effect of accretion and change in the mass of the
planets.\\
Melita \& Woolfson (1996) have performed numerical integrations of
a three-body problem to study the combined effect of resonances
and dissipative forces on the production of stable configurations.
It should be interesting to improve the model, by means of an
appropriate disc model, letting  mass of the planets change by
accretion, and using the formulas for dynamical friction force
used in this paper, instead of the form they took from the stellar
systems evolution theory, and study the effect of different
choices for the initial masses of the planets.
\section*{Acknowledgments}
We are grateful to E. N. Ercan for stimulating discussions during
the period in which this work was performed. A. Del Popolo would
like to thank Bo$\breve{g}azi$\c{c}i University Research
Foundation for the financial support through the project code
01B304.




\newpage
%

\begin{figure}
\caption[]{Evolution of the
surface density. The solid line represents
the disc model of this paper (see section
(2)) while the dashed line represents Stepinski's (1998) result.
In this plot, the disc mass is $0.1 M_{\odot}$, $\alpha=0.01$, $\tau_{\rm
crit}=1.78$ and the angular momentum is $4 \times
10^{52} {\rm g cm^2/s }$. The plots, from top to bottom, represent
the surface density at times $t=10^5$, $10^6$, $10^7 {\rm yr}$.}
\label{Fig. 1} \psfig{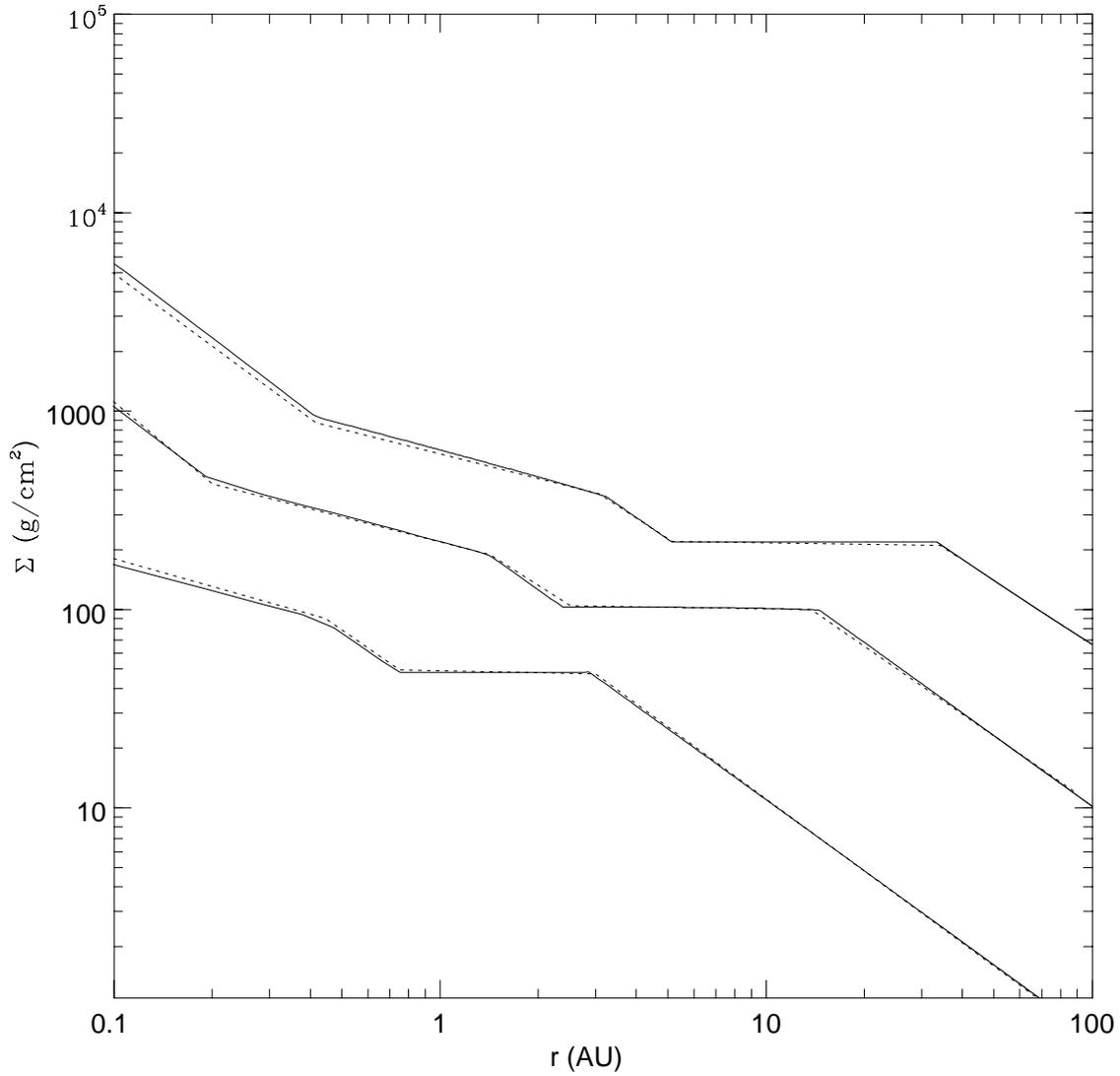}
\end{figure}

\begin{figure}
\caption[]{Comparison of the disc model of this paper, solid line,
with same parameters as Fig. 1, with the
numerical simulations of Reyes-Ruiz \& Stepinski (1995), dashed
line, and $\Sigma$ obtained using our model but only the viscosity
regime SIGOR, dotted line.
As in Fig. 1, the plots, from top to bottom, represent
the surface density at times $t=10^5$, $10^6$, $10^7 {\rm yr}$.}
\label{Fig. 2} \psfig{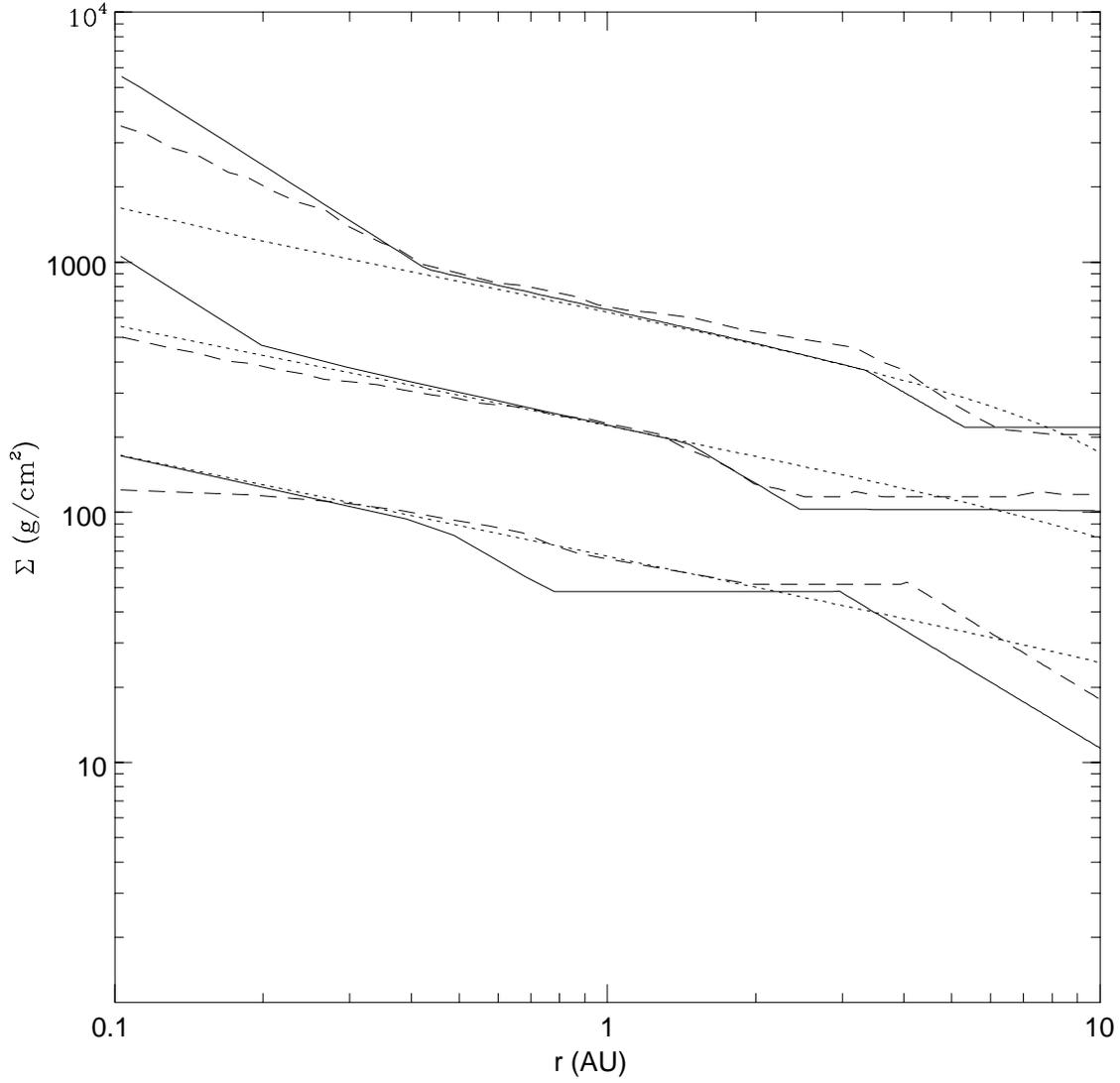}
\end{figure}

\begin{figure}
\caption[]{Radial distribution of temperature, $T(r,t)$, at
selected times ($10^5 {\rm yr}$, $10^6 {\rm yr}$, $10^7 {\rm yr}$) (from
top to bottom)
for a disc of $M_{\rm d}=0.1 M_{\odot}$, $\alpha=0.01$,
$\tau_{\rm crit}=1.78$ and $J_{\rm d}=4 \times
10^{52} {\rm g cm^2/s }$.
The solid line represents $T(r,t)$ obtained with Stepinski's (1998)
model, the dashed line the prediction for temperature of the model
of this paper and the dotted line the approximation given by using
only the SIGOR regime.} \label{Fig. 3} \psfig{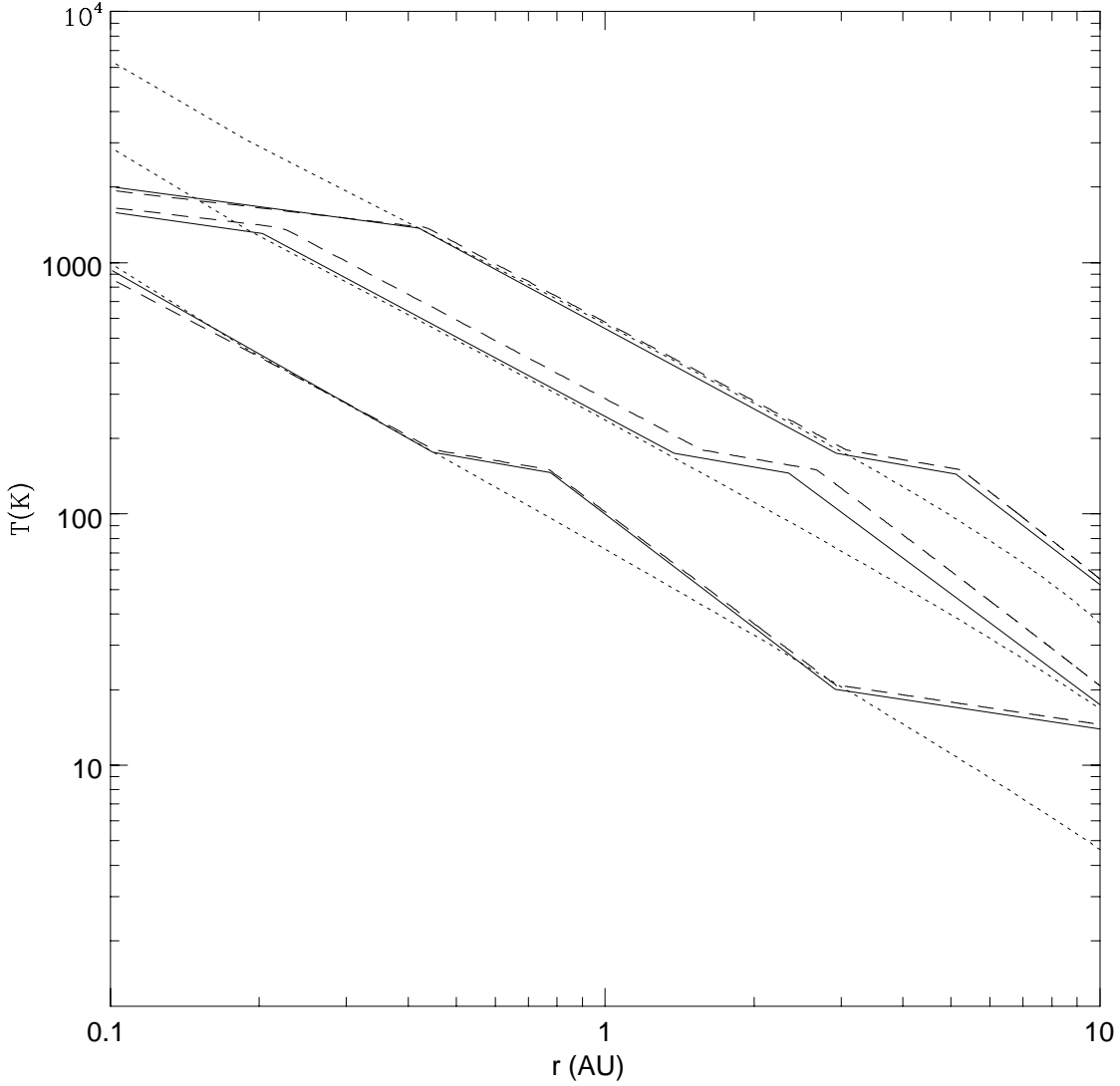}
\end{figure}

\begin{figure}
\caption[]{Evolution of the mass of the nebula
for a disc of $M_{\rm d}=0.245 M_{\odot}$, $\alpha=0.01$.
The solid line represents the surface density of the gas $\times 0.01$,
while the dashed line the evolution of long lived particles (see
Stepinski \& Valageas 1996) of $10^4 {cm}$.}
\psfig{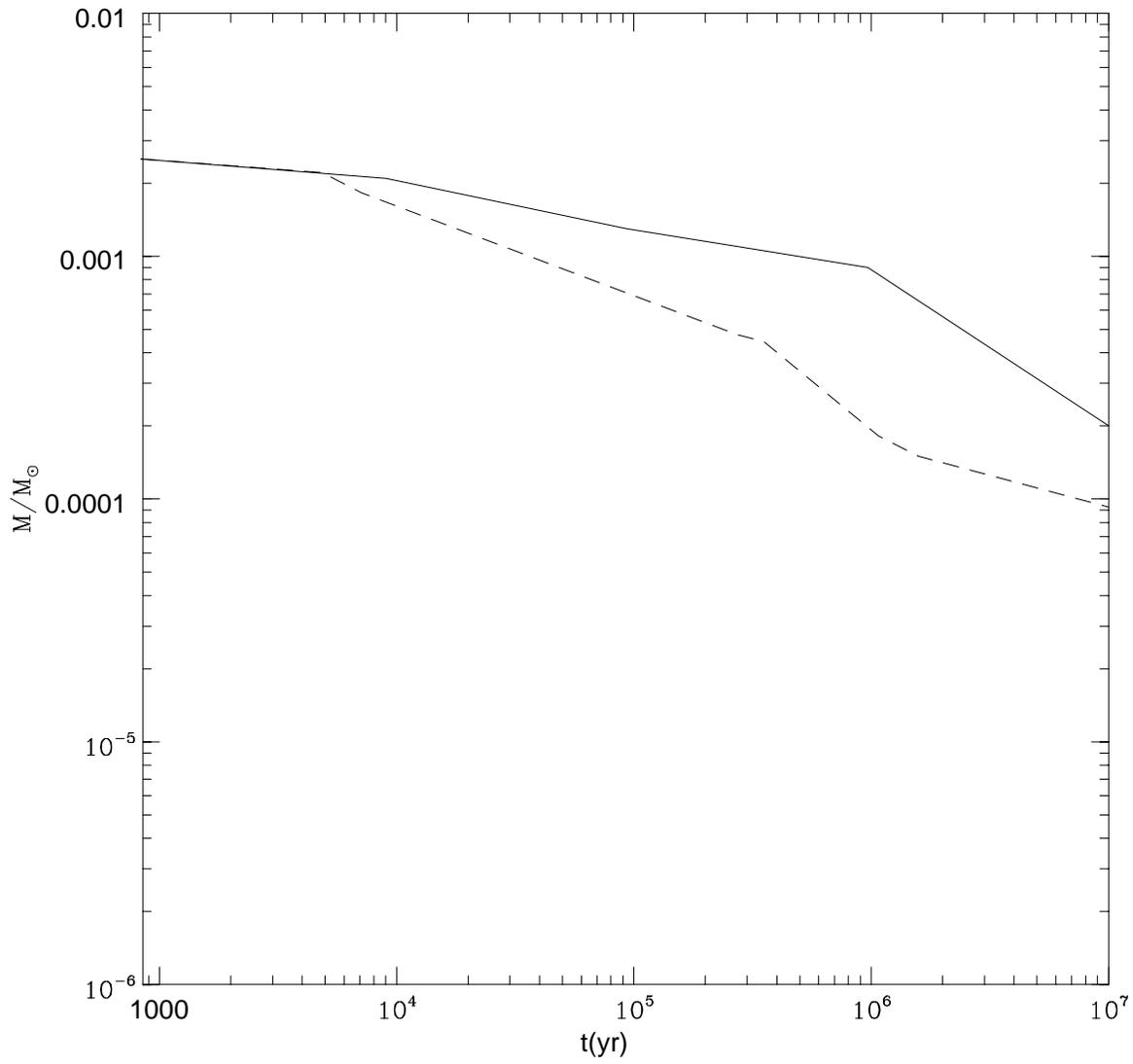}
\end{figure}

\begin{figure}
\label{Fig. 1} \centerline{\hbox{ \psfig{figure=mb738.f5,width=9.5cm} (a)
\psfig{figure=mb738.f6,width=9.5cm} (b)
}}
\end{figure}
\begin{figure}
\centerline{\hbox{ \psfig{figure=mb738.f7,width=9.5cm} (c)
\psfig{figure=mb738.f8,width=9.5cm} (d)
     }}
\caption[]{Evolution of the mass of the nebula
for a disc of $M_{\rm d}=0.023 M_{\odot}$, for $\alpha=0.1$ (a), $\alpha=0.01$ (b), $\alpha=0.001$ (c), $\alpha=0.0001$ (d).
The dashed line represents the surface density of the gas $\times 0.01$,
while the dot-dashed line the evolution of solids.}
\end{figure}

\begin{figure}
\caption[]{Evolution of
semi-major axis of a 1 $M_{J}$ planet in a disc with planetesimals
whose surface
density is 1\% of the gas and having $\alpha=0.001$.
The values of the initial disc mass are
$M_{\rm D}$: 0.1, 0.01, 0.005, 0.001, 0.0005, 0.0001 $M_{\odot}$,
for the solid line, dotted line, short-dashed line, long-dashed
line, dot-short dashed line, dot-long dashed line, respectively.}
\label{Fig. 4} \psfig{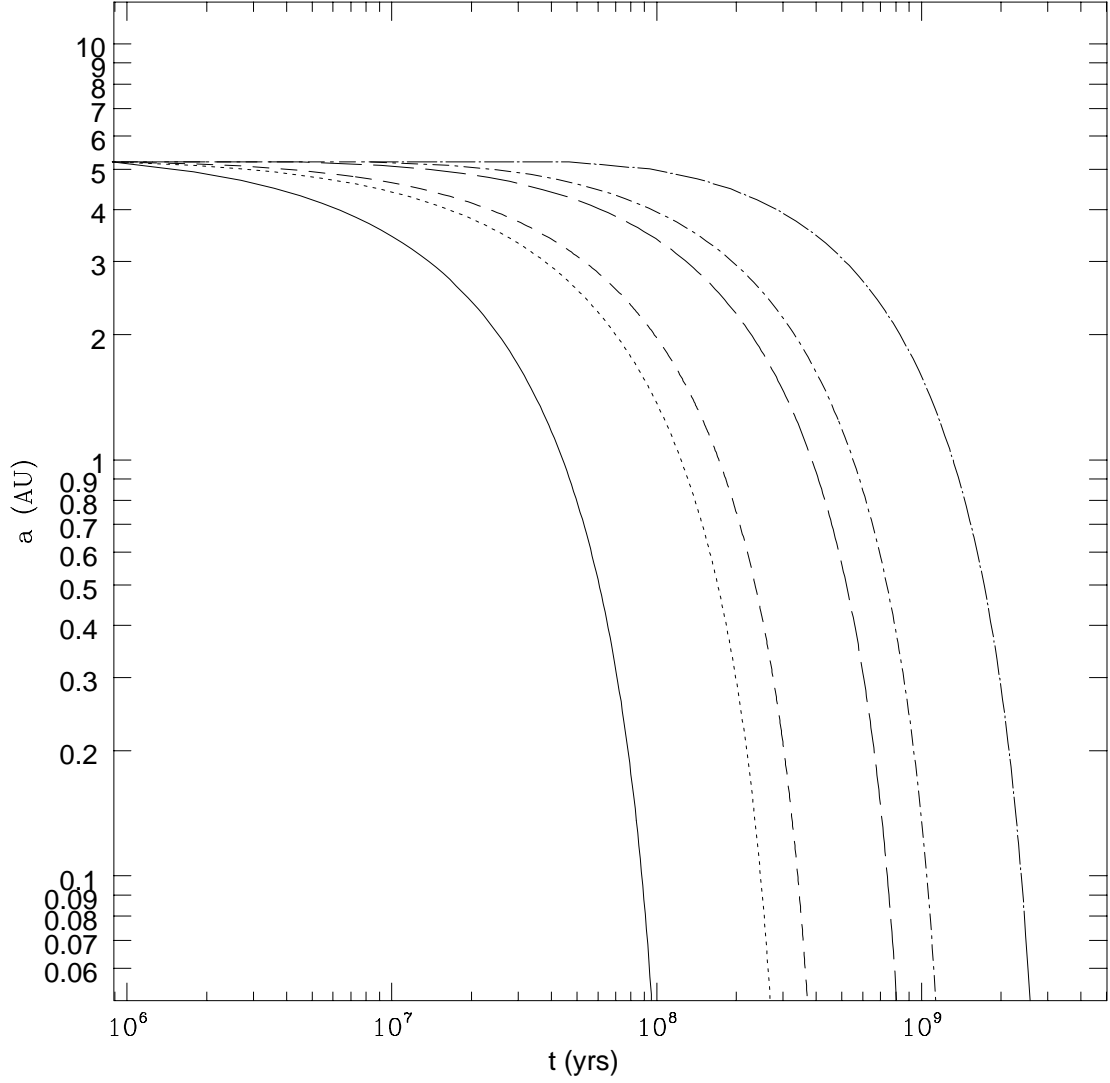}
\end{figure}

\begin{figure}
\caption[]{Same as Fig. 4 but now $\alpha=0.01$.}
\label{Fig. 5} \psfig{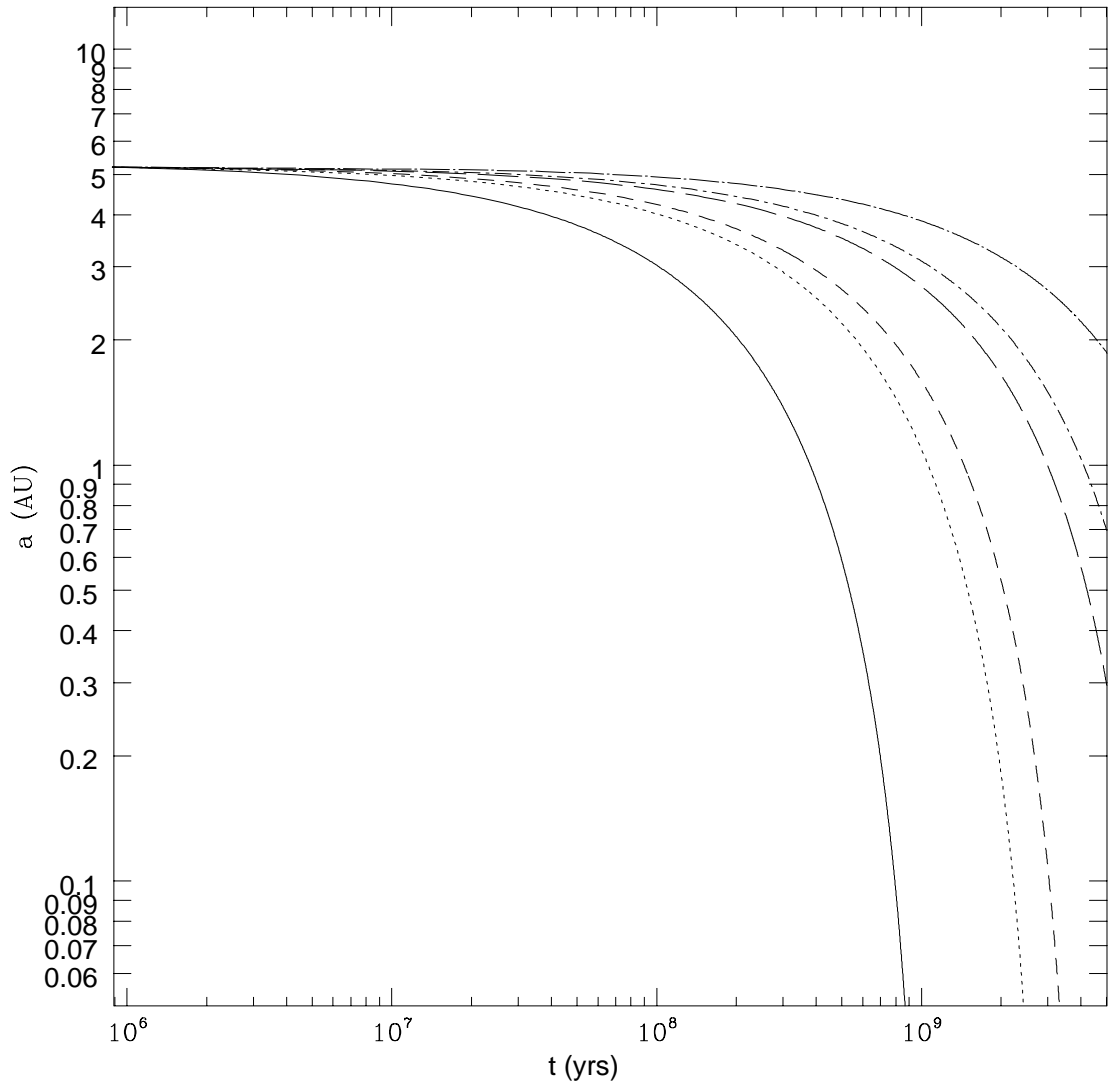}
\end{figure}

\begin{figure}
\caption[]{Same as Fig. 5 but now $\alpha=0.1$.}
\label{Fig. 6} \psfig{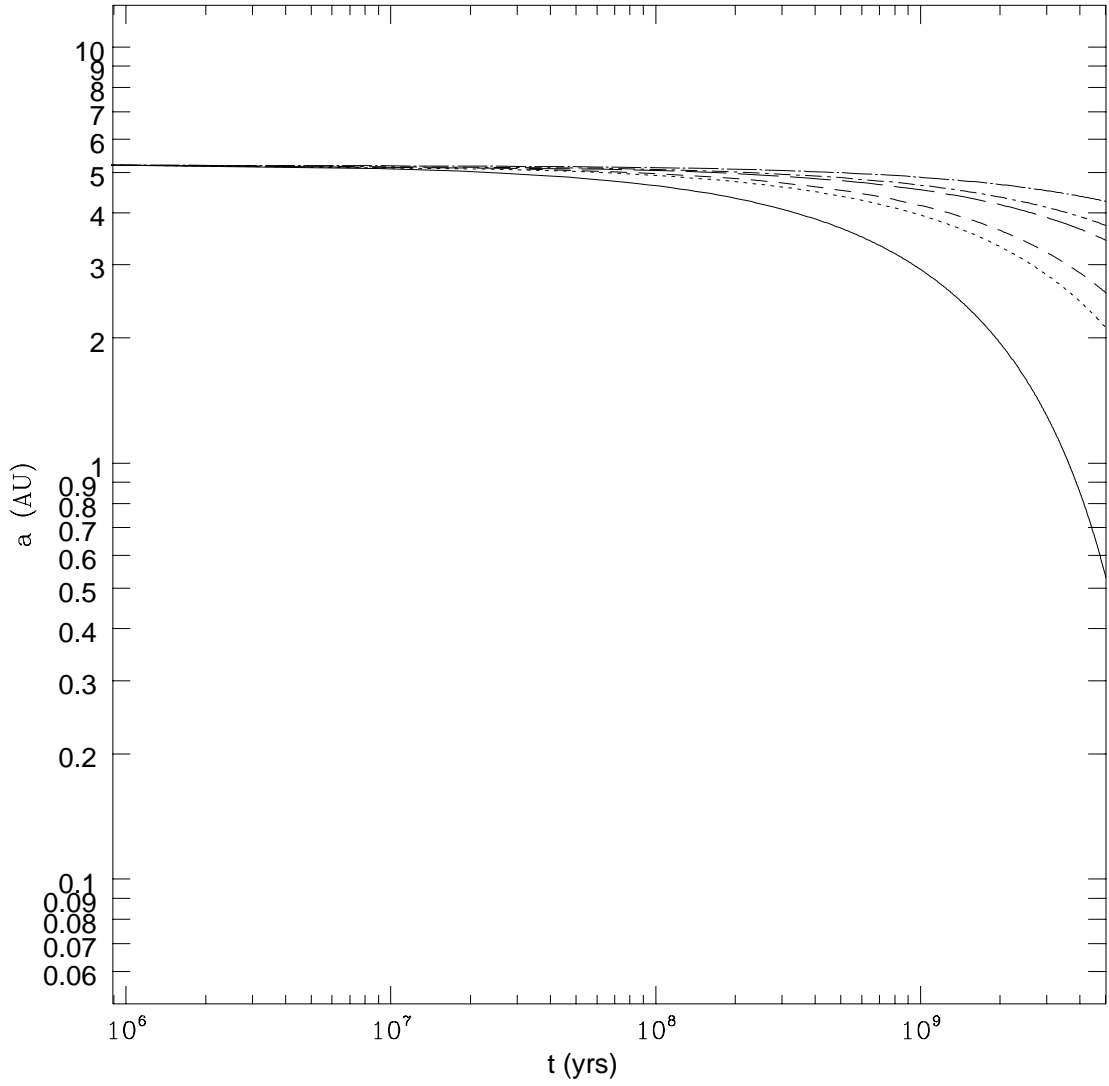}
\end{figure}

\end{document}